\documentstyle[aps]{revtex}
\begin{document}
\draft
\title{
A Phenomenological Description 
of the Non-Fermi-Liquid Phase of MnSi
\footnote{This Short Note will be published in J. Phys. Soc. Jpn. 
Vol. 73 (2004) No. 11. }
}
\author{
O. Narikiyo
}
\address{
Department of Physics, 
Kyushu University, 
Fukuoka 810-8560, 
Japan
}

\vskip 10pt

\date{
April 28, 2004
}

\vskip 10pt

\maketitle
\begin{abstract}
  In order to understand the non-Fermi-liquid behavior 
of MnSi under pressure 
we propose a scenario on the basis of the multispiral state 
of the magnetic moment. 
  This state can describe the recent critical experiment 
of the Bragg sphere in the neutron scattering 
which is the key ingredient of the non-Fermi-liquid behavior. 
\end{abstract}
\vskip 30pt

  MnSi is one of the most extensively studied materials 
from the viewpoint of itinerant magnetism. 
  Recently its non-Fermi-liquid (NFL)
nature~\cite{Nature1,Nature2,Nature3} 
which can not be understood 
within the standard theory~\cite{Hertz,Moriya,Millis} 
has been reported. 
  We will give a phenomenological description 
of this NFL behavior in this Short Note 
as a starting point for constructing a microscopic theory. 

  MnSi is a helimagnetic metal at ambient pressure 
for the temperature $T$ 
below the transition temperature $T_{\rm c}=29.5$K. 
  $T_{\rm c}$ is suppressed 
by the application of hydrostatic pressure $p$ 
and vanishes at the critical pressure $p_{\rm c}=14.6$kbar. 
  For $T>T_{\rm c}$ MnSi is a paramagnetic metal. 
  The phase transition changes its character 
at $p=12{\rm kbar}\equiv p^*$. 
  For $p<p^*$ the transition is of second order, 
while for $p>p^*$ it is of weakly first order. 
  It should be noted that a crossover line $T=T_0(p)$, 
which merges to the transition line $T=T_{\rm c}(p)$ at $p=p^*$, 
exists in $p-T$ plane. 
  For $T_{\rm c}<T<T_0$ the local magnetic moment exists 
even in the paramagnetic phase. 
  Here $T_{\rm c}=0$ for $p>p_{\rm c}$. 
  These behaviors in $p-T$ plane are summarized, 
for example, in Fig.~1 of ref.~3. 

  The NFL behavior in the DC resistivity~\cite{Nature1,Nature2} 
is observed roughly for $T_{\rm c}<T<T_0$ 
as stated in the caption of Fig.~3 in ref.~3. 
  In this region 
the resistivity $R$ is proportional to $T^{3/2}$. 
  For $p>p_{\rm c}$ where $T_{\rm c}=0$ 
this fractional power dependence persists down to a few mK. 
  This behavior is unexpected in two points. 
  Firstly 
the transition at $T_{\rm c}$ is of first order for $p^*<p<p_{\rm c}$, 
while such a fractional power dependence is expected 
near the second order transition. 
  Secondly there is no characteristic scale of temperature 
in the resistivity for $p>p_{\rm c}$, 
while such a vanishing of the scale is expected 
only at the critical pressure of the second order transition. 

  In order to understand the NFL behavior 
a critical experiment by neutron scattering 
has been reported very recently.~\cite{Nature3} 
  Across the first order phase transition line in $p-T$ plane 
the unlocking of the spiral occurs. 
  Namely, it is observed that 
the scattering intensity in the paramagnetic phase 
spreads around the Bragg sphere 
of radius $Q \sim 0.043$\AA$^{-1}$ in reciprocal lattice space, 
while in the long-range ordered helimagnetic phase at ambient pressure 
the magnetic Bragg peak is observed 
at the point of $\vec Q = 0.037$\AA$^{-1}(1,1,1)$ 
which specifies the spiral of the magnetic moment in real space. 

  On the basis of this neutron scattering experiment 
we try to explain the NFL behavior 
in the temperature dependence of the resistivity 
in a phenomenological manner. 
  Firstly we discuss the nature of the transition. 
  For $p<p^*$ the transition at $T=T_{\rm c}$ 
is a simple spin-density wave formation. 
  The local moment is absent for $T>T_{\rm c}$ in this pressure range 
and the transition is of second order. 
  For $p>p^*$ 
non-vanishing global order parameter is attained only for $T<T_{\rm c}$, 
while the local moment is formed for $T<T_0$. 
  In the region $T_{\rm c}<T<T_0$ 
a quasi-long-range correlation is developed, 
although the global order parameter vanishes. 
  Since the local moment is already formed for $T>T_{\rm c}$, 
the transition at $T=T_{\rm c}$ becomes first order in this pressure range. 
  The driving force for the global order is weak 
and the transition is of weakly first order. 

  As an candidate for the driving force 
we can think of the nesting of the Fermi surface, 
since such a weak global effect 
arises from the low-energy states around the Fermi energy. 
  Some nesting tendency of the Fermi surface 
is seen in the band calculation.~\cite{band,JP} 
  It is naturally expected 
that the preferred direction of the spiral 
is determined by the nesting vector. 
  In the magnetically ordered state 
the Fermi surface is gapped around the Bragg wave-vector. 
  The wave vector is determined by the nesting of the Fermi surface. 
  The gap is formed to gain the potential energy 
in spite of the kinetic energy loss. 
  By the application of hydrostatic pressure 
the relative importance of the kinetic energy increases 
and the gap collapses at $T=T_{\rm c}$. 
  Across the first-order transition for $p>p^*$ 
the Bragg peak in the neutron scattering intensity 
spreads over the Bragg sphere. 
  This unlocking of the spiral occurs 
in order to gain kinetic energy. 
  The dissociation of $T_{\rm c}$ and $T_0$ for $p>p^*$ 
results from the fluctuation of the magnetic moments 
due to the kinetic energy. 

  It should be noted that the absence of inversion symmetry in MnSi 
related to B20 crystal structure is one of the most important features 
of this material. 
  The helimagnetism is discussed to be derived 
from the Dzyaloshinsky-Moriya interaction allowed for this symmetry 
from the view point of localized magnetic moments.~\cite{NYHK,BJ} 
  On the other hand, MnSi is well described 
as an itinerant magnet~\cite{Moriya,Fujimori} 
so that the same effect is realized in the renormalized band structure 
for quasiparticles. 
  Namely, the electronic state of MnSi is described as a Fermi liquid 
with a quasiparticle band 
into which the effect of localized magnetic moments 
are renormalized.~\cite{Fujimori} 
  The renormalized quasiparticle band has nesting tendency.~\cite{band,JP} 
  Since we are interested in the low-temperature behavior 
around the quantum critical point, 
the approach on the basis of the Fermi liquid is more appropriate 
than that on the basis of the localized magnetic moments. 

  Since a formation of the Bragg circle 
has been theoretically reported in a two-dimensional case 
on the basis of a double spiral state~\cite{double1,double2} 
where the spiral of the magnetic moment is specified by two basis vectors, 
we expect that the Bragg sphere is derived 
from some multispiral state. 
  The microscopic derivation of the multispiral state 
needs detailed information on the Fermi surface 
and should be done in a future study. 
  The absence of the characteristic temperature 
even in the region away from the quantum critical point 
can be understood by the mechanism of the gradual change of the wave vector 
specifying the multispiral state, 
while the nesting vector specifying the Bragg peak is uniquely defined. 
  Thus in the wide range for $p>p_{\rm c}$ 
the coherence length becomes divergently large on the Bragg sphere 
in the limit of $T \rightarrow 0$. 
  In this paper we use the word, the coherence length, after ref.~3 
and it should be distinguished from the true correlation length 
as discussed below. 
  The microscopic derivation of the above mechanism 
on the basis of the gradual change of the Fermi-surface shape 
should be also done in a future study. 

  Next we discuss the NFL nature 
observed in the temperature dependence of the resistivity. 
  In the wide region of interest, $p>p_{\rm c}$, 
the coherence length can be assumed to be divergently large~\cite{Nature3} 
in the limit of $T \rightarrow 0$. 
  Thus the coherence length $\xi$ is approximately modeled as 
\begin{equation}
\xi^{-2} \sim \alpha T,
\label{xi}
\end{equation}
where the right-hand-side term is induced 
by the mode-mode coupling~\cite{Moriya,Millis} 
and $\alpha$ is a temperature-independent constant. 
  This form is derived for non-critical state, 
while we have approximately set 
$\xi^{-2} \rightarrow 0$ when $T \rightarrow 0$ 
in consistent with the resolution-limited scattering intensity.~\cite{Nature3} 
  With this $\xi$ the imaginary part of 
the dynamical spin susceptibility at low energy 
can be modeled as 
\begin{equation}
\chi''({\vec q},\omega) 
= { D \omega / \Gamma \over 
    [ \xi^{-2} + A({\vec q}-{\vec Q})^2 ]^2 + (\omega / \Gamma)^2 },
\label{chi}
\end{equation}
where ${\vec Q}=Q(1,1,0)$ with $Q \sim 0.043$\AA$^{-1}$. 
  $A$ and $D$ are constants 
independent of ${\vec q}$, $\omega$ and $T$. 
  Here we have taken into account the fact 
that the notion of the Bragg sphere is an approximate one and   
that the scattering intensity is strongest at this ${\vec Q}$ 
as observed in the experiment.~\cite{Nature3} 
  The parameter for the dissipation $\Gamma$ is 
of the order of $v_{\rm F}Q$ 
where $v_{\rm F}$ is the renormalized Fermi velocity for quasiparticles. 
  This form is valid at low energy of $\omega < \Gamma$ 
and sufficient for the discussion 
of the low-temperature behavior of the resistivity. 

  It should be noted that the coherence length $\xi$ 
used in this paper is effective only for the modeling 
of the imaginary part of the susceptibility 
at low energy of $\omega < \Gamma$. 
  Actually the divergent behavior 
in the neutron scattering intensity~\cite{Nature3} 
is related to the imaginary part of the susceptibility 
in the low-energy limit. 
  The true correlation length or the thermodynamic susceptibility 
is determined by the real part, $\chi(\vec q)$ at $\omega = 0$, 
which is related to the imaginary part by the Kramers-Kronig relation as 
$ \chi(\vec q) = {1 \over \pi} \int_{-\infty}^\infty {\rm d}\omega 
                  \chi''({\vec q},\omega) / \omega $. 
  The behavior of the true correlation length is determined 
not only by low-energy process but also by high-energy process 
and does not necessarily coincide with the coherence length 
characterizing the low-energy behavior. 
  Such a situation is similar 
to the case of cuprate superconductors.~\cite{NM} 

  Using this $\chi''({\vec q},\omega)$ 
of $z=2$ form, where $z$ is the dynamical exponent, 
we calculate the resistivity $R$. 
  It is easily evaluated for our non-critical state 
as a correction to the Fermi-liquid behavior~\cite{KY} as 
\begin{equation}
R \propto T^2 \sum_{\vec q}
{ 1 \over [ \xi^{-2} + A({\vec q}-{\vec Q})^2 ]^2 }.
\label{R}
\end{equation}
  Here the extracted factor $T^2$ represents the Fermi-liquid behavior 
and is consisitent with the form of the dissipative term $\omega/\Gamma$ 
in $\chi''({\vec q},\omega)$. 
  The summation in the right-hand side representing the correction 
is proportional to the integral 
\begin{equation}
\int_0^{p_{\rm c}}
  { p^2{\rm d}p \over ( p^2 + \xi^{-2} )^2 }
\sim \xi
\sim (\alpha T)^{-1/2},
\label{int}
\end{equation}
where we have set as ${\vec p}={\sqrt A}({\vec q}-{\vec Q})$. 
  Thus we obtain 
\begin{equation}
R \propto T^{3/2},
\label{RT}
\end{equation}
in the low-temperature limit. 
  This form of resistivity scaling is valid 
for $T < \Gamma$ 
where the assumed form of $\chi''({\vec q},\omega)$ is dominant 
in the evaluation of the resistivity. 
  Since $v_{\rm F}$ and $Q$ depend only weakly on the applied pressure, 
the crossover temperature ${\tilde T} \equiv \Gamma$ 
also depends only weakly on the pressure 
in consistent with the experiment.~\cite{Nature1,Nature2,Nature3} 

  In this Short Note 
we have proposed the multispiral state 
in order to explain the key experiment of 
the Bragg sphere in the neutron scattering. 
  A microscopic description of such a state 
needs detailed information on the Fermi surface 
and is left as a future study. 
  On the basis of the quasi-long-range correlation 
observed as the Bragg sphere 
the NFL behavior in the temperature dependence 
of the resistivity is naturally understood. 

\vskip 30pt


\end{document}